\documentclass[%
 reprint,
superscriptaddress,
 amsmath,amssymb,
pra,
]{revtex4-1}

\usepackage{graphicx}
\usepackage{dcolumn}
\usepackage{bm}

\usepackage[dvipdfmx]{hyperref}
\usepackage{mathrsfs}
\usepackage[usenames]{color}
\def\Tr{\operatorname{Tr}}

\def\im{\operatorname{Im}}
\def\arctanh{\operatorname{arctanh}}
\def\arccosh{\operatorname{arccosh}}
\def\abs#1{\lvert #1 \rvert}
\def\Chi{X}

\begin{document}

\preprint{APS/123-QED}

\title{Collective excitation and stability of flow-induced gapless Fermi superfluids}
\author{Hiroki Yamamura}
\email{yamamura@kh.phys.waseda.ac.jp}
\affiliation{Department of Physics, Waseda University, Shinjuku, Tokyo 169-8555, Japan}
\author{Daisuke Yamamoto}
\email{d-yamamoto@riken.jp}
\affiliation{Condensed Matter Theory Laboratory, RIKEN, Wako, Saitama 351-0198, Japan}

\date{\today}

\begin{abstract}
We study the collective excitation and stability of superfluid Fermi gases flowing with a constant velocity in three-dimensional free space. In particular, we investigate a possible gapless superfluid state induced by the superflow using the mean-field theory and the generalized random-phase approximation (GRPA). For weak attractive interactions, we find that the mean-field superfluid order parameter can take a nonzero value even after the superflow velocity exceeds the threshold for the onset of Bogoliubov quasiparticle excitations. Since the Cooper pairs are only partially broken by the quasiparticle excitations, a gapless superfluid state can be formed over a certain range of superflow velocity above the pair-breaking onset. In addition to the usual quasiparticle-pair continuum and the Anderson-Bogoliubov collective mode, the GRPA excitation spectrum of the gapless superfluid state has a quasiparticle-quasihole continuum and a second collective mode. We find that the long-wavelength excitations of the second collective mode eventually cause dynamical instability of the system when the superflow velocity increases. However, the gapless superfluid state still remains stable in a narrow but finite range of superflow velocity.

\begin{description}
\item[PACS numbers]
03.75.Kk, 03.75.Ss, 67.85.De
\end{description}
\end{abstract}

\pacs{Valid PACS appear here}
\maketitle


\section{\label{sec1}introduction}

Since the experimental realization of Bose-Einstein condensation (BEC) in a weakly interacting Bose gas~\cite{anderson.sci.95,davis.prl.95}, ultracold atomic systems have provided an ideal testing ground for studying the superfluid properties of neutral gases. Especially, ultracold two-component Fermi gases have attracted great interest in the expectation that highly controllable systems are provided for simulating strongly correlated electron phenomena, such as high-temperature superconductivity~\cite{chen.prep.05}. A key advantage of the atomic Fermi systems is that the inter-component interaction can be widely tuned by the use of the Feshbach resonance technique~\cite{chin.rmp.10}. Taking the advantage, the smooth crossover from the Bardeen-Cooper-Schrieffer (BCS) superfluid of weakly coupled Cooper pairs to the BEC of tightly bound molecules has been successfully demonstrated~\cite{regal.prl.04,zwierlein.prl.04,bartenstein.prl.04,kinast.prl.04,zwierlein.sci.05}. The BCS-BEC crossover~\cite{eagles.pr.69,leggett.ln.80,nozieres.jltp.85,sademelo.prl.93,engelbrecht.prb.97,randeria.tb.95,giorgini.rmp.08} is ubiquitous in various fields of physics such as quark matter~\cite{alford.rmp.08} and neutron stars~\cite{alford.rmp.08,chamel.lrr.08,margueron.prc.07,gezerlis.prc.08} as well as solid-state materials~\cite{chen.prep.05}.

One of the most characteristic signatures of superfluidity is dissipationless superflow, which can be stable up to a certain critical velocity. In a superfluid atomic gas, the critical velocity can be directly measured by stirring the gas with a laser beam~\cite{raman.prl.99,onofrio.prl.00}, by applying a sudden displacement of the confinement potential~\cite{cataliotti.sci.01,fertig.prl.05} or by the use of a moving optical lattice~\cite{miller.prl.07,mun.prl.07,fallani.prl.04}. Therefore, the superflow properties and the critical velocity have been studied with great interest both experimentally~\cite{cataliotti.sci.01,fertig.prl.05,miller.prl.07,mun.prl.07,fallani.prl.04,ryu.prl.07,desbuquois.np.12} and theoretically~\cite{combescot.pra.06,spuntarelli.prl.07,sensarma.prl.06,burkov.prl.08,ganesh.pra.09,yunomae.pra.09,tsuchiya.pra.12,yamamoto.prl.13,watanabe.pra.09,watanabe.pra.11,altman.prl.05,iigaya.pra.06,wilson.prl.10,danshita.pra.10,saito.pra.12,kunimi.prb.12} in cold atomic systems. A criterion for the occurrence of dissipationless superflow was first proposed by Landau~\cite{landau.jpussr.41} in the context of superfluid $^4$He. According to the Landau criterion, an elementary excitation can be created only if the velocity of a superflow exceeds the so-called Landau critical velocity~\cite{landau.jpussr.41}
\begin{align}\label{eq:landau criterion}
v_{\mathrm{L}}=\min_p\left[ \dfrac{\varepsilon_p}{p}\right],
\end{align}
where $\varepsilon_p$ is the energy of an elementary excitation of momentum $p$. In uniform superfluids of two-component fermions, it is known that two different kinds of elementary excitation $\varepsilon_p$ can be a cause of the energetic (Landau) instability of superflow depending on the strength of the inter-component attraction~\cite{combescot.pra.06,spuntarelli.prl.07}. In the strong-coupling BEC regime, the instability of superflow is induced by bosonic long-wavelength excitations of the collective sound mode, known as the Anderson-Bogoliubov (AB) mode. On the other hand, in the weak-coupling BCS regime, a fermionic quasiparticle excitation with a finite momentum can be created before the bosonic collective excitations occur. 
As a result, the Landau critical velocity $v_{\mathrm{L}}$ across the BCS-BEC crossover exhibits nonmonotonic behavior with a maximum that is located close to the unitarity limit. At the maximum, the mechanism of the instability changes from fermionic to bosonic excitations~\cite{combescot.pra.06,spuntarelli.prl.07,sensarma.prl.06}. The predicted pronounced peak of the critical velocity near unitarity has been confirmed experimentally in ultracold superfluid Fermi gases with the use of a shallow moving optical lattice~\cite{miller.prl.07}.

The influence of a finite superflow in ultracold atomic gases has been studied also in the presence of deep optical lattices where the tight-binding approximation is valid~\cite{burkov.prl.08,ganesh.pra.09,yunomae.pra.09,tsuchiya.pra.12,yamamoto.prl.13,altman.prl.05,iigaya.pra.06,danshita.pra.10,saito.pra.12}. In the low density limit, the attractive Hubbard model in the presence of superflow exhibits a similar behavior to the homogeneous case~\cite{ganesh.pra.09,yunomae.pra.09,sofo.prb.92} mentioned above. Away from the low density limit, the AB mode possesses a roton-like dip structure at short wavelength due to density-wave fluctuations. Consequently the short-wavelength collective excitation can occur before the other excitations when the superflow velocity increases. The instability mechanism of the lattice systems changes depending on the filling, the attraction strength and the geometry of the lattice~\cite{ganesh.pra.09,yunomae.pra.09,tsuchiya.pra.12}. It has been also discussed that a charge-density-wave order may appear coexisting with superfluidity due to the softening of the short-wavelength collective excitation. For two-dimensional (2D) and three-dimensional (3D) hypercubic optical lattices, such a ``flowing supersolid" state is predicted to be unstable because of the negative superfluid stiffness~\cite{burkov.prl.08}. On the other hand, for a Fermi superfluid in a moving kagome optical lattice~\cite{yamamoto.prl.13}, the particular geometry of the kagome lattice introduces anisotropy in the hopping amplitude of Cooper pairs, leading to a density modulation without breaking the superfluidity.

Let us revisit the subject of the flowing homogeneous Fermi superfluids in 3D free space. As stated above, according to the Landau criterion, the superflow is destabilized by bosonic long-wavelength excitations in the strong-coupling BEC regime, while fermionic quasiparticle excitations occur in the weak-coupling BCS regime~\cite{combescot.pra.06}. However the latter does not mean the breakdown of superfluidity since the fermionic excitations cannot lower the energy of the system indefinitely due to the Pauli exclusion principle \cite{ganesh.pra.09,bardeen.rmp.62,zagoskin.tb.98,wei.prb.09}. In other words, the Landau instability due to the fermionic excitations just correspond to the onset for Cooper pair breakings. In fact, it is known that in 3D superconductors, a gapless superconducting state that possesses nonzero order parameter but zero energy gap~\cite{abrikosov.zetf.61} can be allowed to appear in a certain range of supercurrent velocity even after pair breaking starts to occur~\cite{ganesh.pra.09,bardeen.rmp.62,zagoskin.tb.98,wei.prb.09}. By analogy with the superconductors, it is necessary to take into account the possibility of a gapless superfluid state that may exist above the Landau critical velocity.

Therefore, in this paper, we study the influence of superflow on Fermi superfluids at zero temperature ($T=0$) considering the possible flow-induced gapless superfluid states. First, we calculate the superflow velocity dependence of the superfluid order parameter within the mean-field approximation. In the BCS regime, we find that a gapless superfluid state can survive even after the superflow velocity exceeds the onset for Cooper pair breaking as in the case of the 3D superconductors. The flow-induced gapless superfluid appears in a considerable range of the superflow velocity for the intermediate attractive interaction (but on the BCS side). However we should note that in a neutral superfluid, the collective AB mode plays a crucial role in the stability of the system in contrast of superconductors, in which the AB mode is pushed up to the plasma frequency due to the long-range Coulomb interaction~\cite{anderson.pr.58}. Therefore, we analyze the collective excitation spectra by employing the generalized random-phase approximation (GRPA)~\cite{engelbrecht.prb.97,combescot.pra.06,cote.prb.93,yunomae.pra.09,minguzzi.epjd.01,bruun.prl.01,ohashi.jpsj.97,ohashi.pra.03,taylor.pra.06,zhang.pra.11,hu.fp.12}. The GRPA excitation spectrum has a continuum of Bogoliubov quasiparticle-pair excitations and the collective AB mode when the superflow is absence~\cite{combescot.pra.06}. In the BCS regime, if the superflow velocity exceeds the onset value for the pair breaking, an additional quasiparticle-quasihole continuum and a second collective mode appear. We find that the excitation energy of the second collective mode can have a nonzero imaginary part even if the superfluid order parameter remains finite. This excitation leads to an exponential growth of arbitrarily small perturbations in time, called dynamical instability~\cite{wu.pra.01,burkov.prl.08}. As a result, there remains only a small window of superflow velocity for stable gapless superfluid states.

This paper is organized as follows. In Sec.~\ref{sec2}, we introduce the order parameter that describes Cooper pairing with nonzero center-of-mass momentum in order to consider the presence of superflow. Within the mean-field approximation, we solve a set of self-consistent equations to determine the region where the gapless superfluid state can appear in the plane of interaction strength and superflow velocity. In Sec.~\ref{sec3}, based on the GRPA, we discuss the stability of the flow-induced gapless superfluid state and present the stability phase diagram of Fermi superfluids in the presence of superflow. Finally, in Sec.~\ref{sec4}, we give a summary and make some remarks on gapless superfluid states in other systems. Throughout the paper, the Planck constant $\hslash $ and the Boltzmann constant $k_{\mathrm{B}}$ are set to be unity $(\hslash=k_{\mathrm{B}}=1)$.

\section{\label{sec2}mean-field analysis}

We describe a system of homogeneous two-component Fermi gases with equal mass and equal population using the following Hamiltonian:
\begin{align}\label{eq:hamiltonian}
\hat{H}=\sum_{\sigma}\int d\mathbf{r}\,&\hat{\psi}_{\sigma}^\dagger (\mathbf{r})\left(-\dfrac{\nabla ^2}{2m}-\mu \right)\hat{\psi}_{\sigma}(\mathbf{r})\nonumber\\
&+\,g\int d\mathbf{r}\,\hat{\psi}_{\uparrow}^\dagger (\mathbf{r})\hat{\psi}_{\downarrow}^\dagger (\mathbf{r})\hat{\psi}_{\downarrow} (\mathbf{r})\hat{\psi}_{\uparrow} (\mathbf{r}),
\end{align}
where $\hat{\psi}_{\sigma}^\dagger (\mathbf{r})$ is the field operator that creates a fermion with pseudospin $\sigma=\uparrow, \downarrow$, $m$ is the mass of the fermion, and $\mu$ is the chemical potential. We perform a mean-field analysis of the Hamiltonian (\ref{eq:hamiltonian}) under the presence of superflow. We assume an attractive $s$-wave interaction ($g<0$) between the ``spin-up" and ``spin-down" particles, which can be tuned by the Feshbach resonance technique~\cite{chin.rmp.10}. For a standard (gapped) superfluid, the effects of superflow can be discussed only by applying the Galilean transformation to the system~\cite{landau.jpussr.41}. However one has to explicitly assume that Cooper pairs possess a nonzero center-of-mass momentum $2m\mathbf{v}$ in order to discuss a possible gapless superfluid state in the presence of superflow with a constant velocity $\mathbf{v}$. We therefore introduce the superfluid order parameter $\Delta$ as
\begin{align*}
\Delta =-\dfrac{g}{V}\sum_{\mathbf{k}} \langle \hat{c}_{-\mathbf{k}+m\mathbf{v},\downarrow}\hat{c}_{\mathbf{k}+m\mathbf{v},\uparrow}\rangle,
\end{align*}
where $V$ denotes the volume of the system and $\hat{c}_{\mathbf{k}\sigma }$ is the Fourier transform of the annihilation operator $\hat{\psi}_{\sigma } (\mathbf{r})$:
\begin{align*}
\hat{\psi }_{\sigma }(\mathbf{r})=\dfrac{1}{\sqrt{V}}\sum_{\mathbf{k}}e^{i\mathbf{k}\cdot \mathbf{r}}\hat{c}_{\mathbf{k}\sigma }.
\end{align*}
In the rest of this section, we derive the velocity $\mathbf{v}$ dependence of the superfluid order parameter $\Delta $, and show that a gapless superfluid state can be formed in a certain range of $\abs{\mathbf{v}}$.

\subsection{\label{subsec:gap}Self-consistent equations}

\begin{figure}
\includegraphics[width=8.4cm]{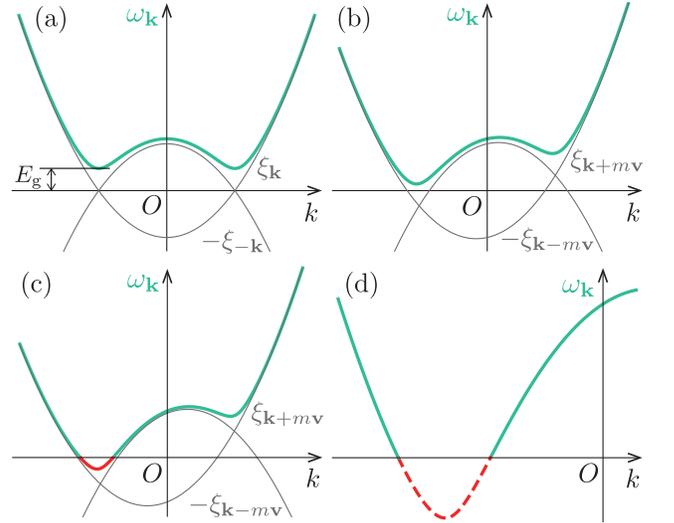}
\caption{\label{fig:quasiparticle dispersions}
(Color online) Schematic plots of the Bogoliubov quasiparticle energy $\omega_{\mathbf{k}}$ (thick solid lines) for (a) $v=0$, (b) $0<v<v_{\mathrm{pb}}$, (c) $v_{\mathrm{pb}}<v<v_{\mathrm{pb}}^*$. We also show the free-particle and free-hole energies, $\xi_{\mathbf{k}+m\mathbf{v}}$ and $-\xi_{-\mathbf{k}+m\mathbf{v}}$ by thin solid lines. Here, $v_{\mathrm{pb}}^*$ is the velocity at which the order parameter vanishes (see the text). (d) is the enlarged view of the region where the Bogoliubov band has negative energy in (c). The absolute value of $k$ denote $\abs{\mathbf{k}}$, and a positive (negative) value of $k$ means that the quasiparticle momentum $\mathbf{k}$ in the same (opposite) direction as superflow velocity $\mathbf{v}$.}
\end{figure}

Applying the standard mean-field approximation to the Hamiltonian (\ref{eq:hamiltonian}), we obtain
\begin{align}\label{eq:mean-field hamiltonian in nambu-gor'kov formalism}
\hat{H}_{\mathrm{MF}}=&\sum_{\mathbf{k}}
\hat{\Psi }_{\mathbf{k}}^\dagger 
\begin{pmatrix}
\xi_{\mathbf{k}+m\mathbf{v}} & -\Delta \\[2pt]
-\Delta & -\xi _{-\mathbf{k}+m\mathbf{v}}
\end{pmatrix}
\hat{\Psi }_{\mathbf{k}} \nonumber \\
&\hspace{70pt}+\sum_{\mathbf{k}}\xi_{-\mathbf{k}+m\mathbf{v}}-\dfrac{\Delta ^2}{g}V,
\end{align}
in the Fourier space. Here, $\hat{\Psi }_{\mathbf{k}}^\dagger =(\begin{matrix} \hat{c}_{\mathbf{k}+m\mathbf{v},\uparrow}^\dagger & \hat{c}_{-\mathbf{k}+m\mathbf{v},\downarrow} \end{matrix})$ is the Nambu-Gor'kov spinor and $\xi _{\mathbf{k}}=\epsilon_{\mathbf{k}}-\mu$ denotes the kinetic energy ($\epsilon _{\mathbf{k}}=\abs{\mathbf{k}}^2/2m$) measured from the chemical potential. The mean-field Hamiltonian $\hat{H}_{\mathrm{MF}}$ can be diagonalized by the following Bogoliubov transformation:
\begin{align}\label{eq:bogoliubov transformation}
\hat{\Psi }_{\mathbf{k}}=
\begin{pmatrix}
u_{\mathbf{k}} & v_{\mathbf{k}}  \\[2pt]
-v_{\mathbf{k}}  & u_{\mathbf{k}} 
\end{pmatrix}
\begin{pmatrix}
\hat{\alpha }_{\mathbf{k}0}  \\
\hat{\alpha }_{-\mathbf{k}1}^\dagger 
\end{pmatrix},
\end{align}
where the Bogoliubov coefficients $u_{\mathbf{k}}$ and $v_{\mathbf{k}}$ are chosen so that the  $2\times 2$ matrix in Eq.~(\ref{eq:mean-field hamiltonian in nambu-gor'kov formalism}) can be diagonalized: ${u_{\mathbf{k}}}{}^2=(1+\xi_{\mathbf{k}}^+/E_{\mathbf{k}})/2$ and ${v_{\mathbf{k}}}{}^2=(1-\xi_{\mathbf{k}}^+/E_{\mathbf{k}})/2$ with $\xi_{\mathbf{k}}^+=(\xi_{\mathbf{k}+m\mathbf{v}}+\xi_{\mathbf{k}-m\mathbf{v}})/2$ and $E_{\mathbf{k}}=\sqrt{\xi _{\mathbf{k}}^+{}^2+\Delta^2}$. Here $\hat{\alpha }_{\mathbf{k}\lambda}^\dagger $ is the creation operator of a Bogoliubov quasiparticle with momentum $\mathbf{k}$ and $\lambda =0,1$. The Bogoliubov quasiparticles $\hat{\alpha }_{\mathbf{k}\lambda }$ and $\hat{\alpha }_{\mathbf{k}\lambda }^\dagger $ satisfy the anticommutation relations $\{\hat{\alpha }_{\mathbf{k}\lambda },\hat{\alpha }_{\mathbf{k}'\lambda '}^\dagger \}=\delta_{\mathbf{k}\mathbf{k}'}\delta_{\lambda \lambda '}$ and $\{\hat{\alpha }_{\mathbf{k}\lambda },\hat{\alpha }_{\mathbf{k}'\lambda '}\}=\{\hat{\alpha }_{\mathbf{k}\lambda }^\dagger ,\hat{\alpha }_{\mathbf{k}'\lambda '}^\dagger \}=0$. Using the Bogoliubov quasiparticle operators, the mean-field Hamiltonian (\ref{eq:mean-field hamiltonian in nambu-gor'kov formalism}) can be rewritten in the diagonal form:
\begin{align}\label{eq:bogoliubov transformed mean-field hamiltonian}
\hat{H}_{\mathrm{MF}}=\sum_{\lambda =0,1}\sum_{\mathbf{k}}\omega_{\mathbf{k}} \hat{\alpha }_{\mathbf{k}\lambda } ^\dagger \hat{\alpha }_{\mathbf{k}\lambda }+E_0,
\end{align}
with
\begin{align*}
E_0=-\sum_{\mathbf{k}}(E_{\mathbf{k}}-\xi_{\mathbf{k}}^+)-\dfrac{\Delta ^2}{g}V.
\end{align*}
The Bogoliubov quasiparticle energy is given by
\begin{align}\label{eq:bogoliubov quasiparticle energy}
\omega_{\mathbf{k}}=E_{\mathbf{k}}+\mathbf{k}\cdot \mathbf{v}.
\end{align}
We schematically show the dispersion $\omega_{\mathbf{k}}$ for several values of $v=\abs{\mathbf{v}}$ in Fig.~\ref{fig:quasiparticle dispersions}.
For $v\not =0$, the quasiparticle dispersion is tilted in the direction opposite to the superflow due to the Doppler shift $\mathbf{k}\cdot \mathbf{v}$. With the superflow velocity $v$ increases, the energy gap $E_{\mathrm{g}}$ decreases monotonically and reaches zero at a certain velocity $v=v_{\mathrm{pb}}$. For $v>v_{\mathrm{pb}}$, one finds that quasiparticle states with negative energy appear for a certain range of $\mathbf{k}$ [shown by the red-dashed curve in Fig.~\ref{fig:quasiparticle dispersions}(d)], where Bogoliubov quasiparticles can be spontaneously created. Such quasiparticle creation occurs in pairs since $\hat{\alpha }_{\mathbf{k}0}$ and $\hat{\alpha }_{-\mathbf{k}1}$ are degenerate with the same energy $\omega _{\mathbf{k}}$. This quasiparticle-pair excitation corresponds to the Cooper pair breaking in terms of the original fermion. Hence one can see that $v=v_{\mathrm{pb}}$ is the threshold velocity for the onset of the pair breaking.

The ground state energy $\mathcal{E}_0$ can be obtained from the diagonalized mean-field Hamiltonian (\ref{eq:bogoliubov transformed mean-field hamiltonian}) as
\begin{align}\label{eq:thermodynamic potential}
\mathcal{E}_0=2\sum_{\mathbf{k}}\omega_{\mathbf{k}}\varTheta(-\omega_{\mathbf{k}})+E_0,
\end{align}
where $\varTheta(\epsilon)$ denotes the Heaviside step function defined by $\varTheta(\epsilon)=1$ for $\epsilon \ge 0$ and $\varTheta(\epsilon)=0$ for $\epsilon<0$. The first term represents the contribution of the quasiparticle-pair creation. For $v<v_{\mathrm{pb}}$, the ground state energy is reduced to $\mathcal{E}_0=E_0$ because the quasiparticle energy $\omega_{\mathbf{k}}$ is always positive [Fig.~\ref{fig:quasiparticle dispersions}(a) and (b)]. On the other hand, for $v>v_{\mathrm{pb}}$, the Heaviside step function $\varTheta(-\omega_{\mathbf{k}})$ takes 1 in a certain region of $\mathbf{k}$-space, and the energy of the system $\mathcal{E}_0$ is lowered from $E_0$ due to the spontaneous creation of quasiparticle pairs. As a result, the negative energy states in the two degenerate Bogoliubov bands [indicated by the red-dashed curve in Fig.~\ref{fig:quasiparticle dispersions}(d)] are filled with the quasiparticles.

The quasiparticle excitations cannot be created indefinitely due to the Pauli exclusion principle. Therefore, a gapless superfluid state that has nonzero order parameter but zero energy gap can be stabilized for $v>v_{\mathrm{pb}}$. The partial breaking of Cooper pairs and the formation of the gapless superfluid state can be seen by calculating the superflow velocity $v$ dependence of the order parameter $\Delta $. The stationary condition $\partial \mathcal{E}_0 /\partial \Delta =0$ leads to the following gap equation:
\begin{align}\label{eq:gap equation}
-\dfrac{1}{g}=\dfrac{1}{V}\sum_{\mathbf{k}}\dfrac{1}{2E_{\mathbf{k}}}[1-2\varTheta(-\omega _{\mathbf{k}})].
\end{align}
It is known that the sum in the right-hand side of Eq.~(\ref{eq:gap equation}) has an ultraviolet divergence, which can be removed by introducing a momentum cutoff or by using the relation between the bare coupling constant $g$ and the effective $s$-wave scattering length $a_s$ \cite{randeria.tb.95}:
\begin{align}\label{eq:regularization}
\dfrac{m}{4\pi a_s}=\dfrac{1}{g}+\dfrac{1}{V}\sum_{\mathbf{k}}\dfrac{1}{2\epsilon_{\mathbf{k}}},
\end{align}
We regularize the ultraviolet divergence by eliminating the coupling constant $g$ from Eqs.~(\ref{eq:gap equation}) and (\ref{eq:regularization}). Furthermore, the total number of fermions $N=V{k_{\mathrm{F}}}^3/3\pi ^2$ ($k_{\mathrm{F}}$ being the Fermi wave vector) satisfies the number equation $N=-\partial \mathcal{E}_0 /\partial \mu $, which should be solved together with the gap equation (\ref{eq:gap equation}). Thus, we obtain a set of self-consistent equations
\begin{align}\label{eq:gap equation2}
-\dfrac{m}{4\pi a_s}=\dfrac{1}{V}\sum_{\mathbf{k}}\left[\dfrac{1}{2E_{\mathbf{k}}}[1-2\varTheta(-\omega _{\mathbf{k}})]-\dfrac{1}{2\epsilon_{\mathbf{k}}}\right],
\end{align}
and
\begin{align}\label{eq:number equation}
\dfrac{{k_{\mathrm{F}}}^3}{3\pi ^2}=\dfrac{1}{V}\sum_{\mathbf{k}}\left[1-\dfrac{\xi _{\mathbf{k}}^+}{E_{\mathbf{k}}}[1-2\varTheta(-\omega _{\mathbf{k}})]\right].
\end{align}
The solution of Eqs.~(\ref{eq:gap equation2}) and (\ref{eq:number equation}) provides a reasonable description of the ground-state properties from the weak-coupling to strong-coupling regime~\cite{eagles.pr.69,leggett.ln.80,nozieres.jltp.85,sademelo.prl.93,engelbrecht.prb.97,randeria.tb.95}. In this formalism, the interaction strength is characterized by the dimensionless parameter $1/k_{\mathrm{F}}a_s$: the weak-coupling and strong-coupling limits correspond to $1/k_{\mathrm{F}}a_s=-\infty $ and $\infty $, respectively.

\subsection{\label{subsec:cv}Flow-induced gapless superfluid}

\begin{figure}
\includegraphics[width=8.4cm]{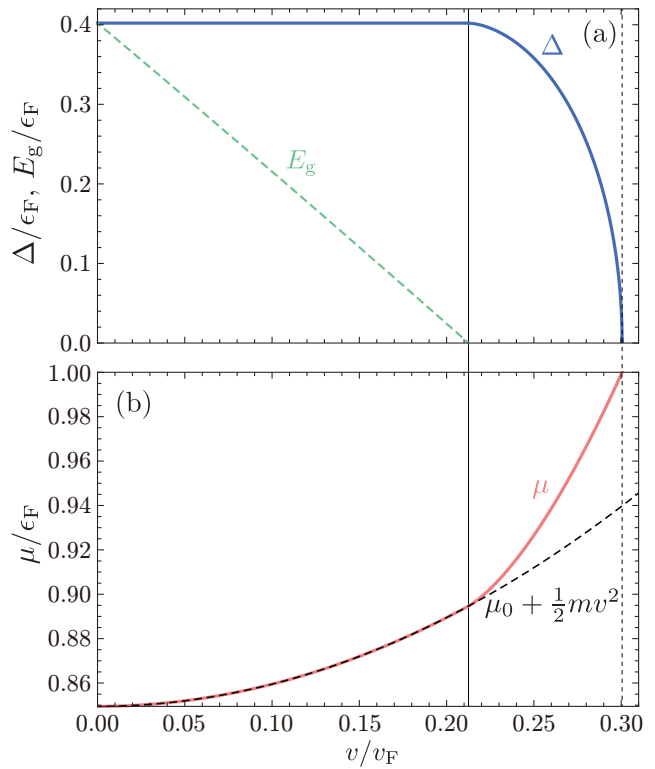}
\caption{\label{fig:self-consistent solutions}
(Color online) The superflow velocity $v$ dependences of (a) the order parameter $\Delta $ (solid line) and the quasiparticle energy gap $E_{\mathrm{g}}$ (dashed line), and (b) the chemical potential $\mu $ (solid line) and $\mu_0+m{v}^2/2$ (dashed line) for $1/k_{\mathrm{F}}a_s=-0.5$. The onset of pair breaking $v_{\mathrm{pb}}$ is indicated by solid vertical line. The dashed vertical line corresponds to the velocity $v_{\mathrm{pb}}^*$ at which the order parameter vanishes.}
\end{figure}

In Fig.~\ref{fig:self-consistent solutions}, we show the numerical solution of the self-consistent equations Eqs.~(\ref{eq:gap equation2}) and (\ref{eq:number equation}) as a function of the superflow velocity $v$  for the interaction parameter $1/k_{\mathrm{F}}a_s=-0.5$.
As we already mentioned above the quasiparticle energy gap $E_{\mathrm{g}}$ decreases monotonically as $v$ increases, and eventually reaches zero at $v=v_{\mathrm{pb}}$. For $0<v<v_{\mathrm{pb}}$, since $\varTheta(-\omega _{\mathbf{k}})=0$ for any value of $\mathbf{k}$, the superflow velocity dependence appears only through $\xi _{\mathbf{k}}^+$ in the self-consistent equations (\ref{eq:gap equation2}) and (\ref{eq:number equation}). Moreover, this effect can be canceled by shifting the chemical potential by $m{v}^2/2$. Therefore the order parameter $\Delta $ takes a constant value $\Delta _0$ for $v< v_{\mathrm{pb}}$ and the chemical potential $\mu $ is simply given by $\mu_0+m{v}^2/2$, where $\Delta _0$ and $\mu _0$ are the solutions of Eqs.~(\ref{eq:gap equation2}) and (\ref{eq:number equation}) for $v=0$. This corresponds to the invariance of the system under the Galilean transformation with a constant velocity $v$. Therefore the onset for the pair breaking $v_{\mathrm{pb}}$ can be obtained by applying the Landau criterion (\ref{eq:landau criterion}) to the excitation energy at $v=0$, $E^0_{\mathbf{k}}=\omega_{\mathbf{k}}|_{v=0}=\sqrt{(\epsilon_{\mathbf{k}}-\mu _0)^2+\Delta _0{}^2}$, as~\cite{combescot.pra.06}
\begin{align}\label{eq:pair-breaking velocity}
v_{\mathrm{pb}}=\min_k\left[\dfrac{E_{\mathbf{k}}^0}{k}\right]=\sqrt{\dfrac{\sqrt{{\Delta_0}^2+{\mu _0}^2}-\mu _0}{m}}.
\end{align}

The main focus of the present paper is that the existence of a gapless superfluid state for $v>v_{\mathrm{pb}}$. As shown in Fig.~\ref{fig:self-consistent solutions}(a), the order parameter remains nonzero even after the superflow velocity $v$ exceeds $v_{\mathrm{pb}}$ and the pair-breaking excitations begin to be created. This is analogous to the case of 3D BCS superconductors with supercurrent~\cite{zagoskin.tb.98}. When the superflow velocity $v$ further increases, the order parameter $\Delta $ decreases monotonically due to the quasiparticle-pair creations, and the chemical potential $\mu $ deviates from $\mu_0 +m{v}^2/2$. Thus, the Galilean symmetry is broken for $v>v_{\mathrm{pb}}$. The order parameter $\Delta $ eventually reaches zero and the chemical potential $\mu $ becomes equal to the Fermi energy $\epsilon_{\mathrm{F}}$ at a certain velocity $v_{\mathrm{pb}}^*$. At $v=v_{\mathrm{pb}}^*$, all Cooper pairs are broken and the system undergoes a second-order transition to the normal fluid. This behavior is similar to the finite temperature case where the order parameter is suppressed by thermally excited quasiparticles and vanishes at a critical temperature. Solving the gap equation (\ref{eq:gap equation2}) with the conditions $\Delta \to 0$ and $\mu=\epsilon_{\mathrm{F}}$, one can find that the offset of the pair breaking $v_{\mathrm{pb}}^*$ is given by
\begin{align}\label{eq:order parameter vanishes}
\theta \tan \theta =\dfrac{\pi}{2k_{\mathrm{F}}a_s}-1,
\end{align}
with $\theta =\arccos(v_{\mathrm{F}}/v_{\mathrm{pb}}^*)$ (see Appendix \ref{appendix} for the details of the derivation).

\begin{figure}
\includegraphics[width=8.4cm]{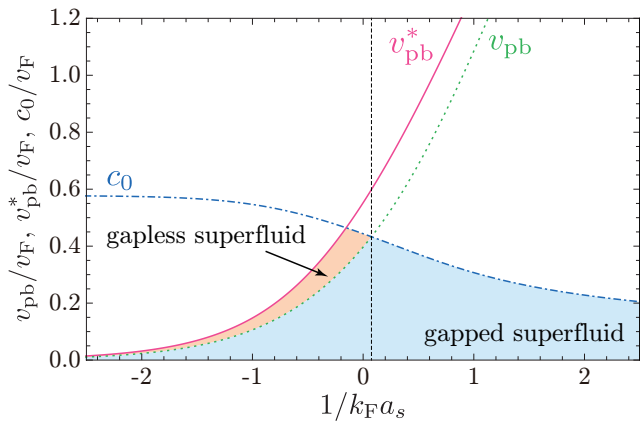}
\caption{\label{fig:mean-field phase diagram}
(Color online) The interaction parameter $1/k_{\mathrm{F}}a_s$ dependences of the onset ($v_{\mathrm{pb}}$; dotted line) and offset ($v_{\mathrm{pb}}^*$; solid line) for the pair breaking and the sound velocity $c_0$ (dot-dashed line). The dashed vertical line corresponds to $(1/k_{\mathrm{F}}a_s)_0$ at which $v_{\mathrm{pb}}=c_0$.}
\end{figure}

We show the $1/k_{\mathrm{F}}a_s$ dependences of $v_{\mathrm{pb}}$ and $v_{\mathrm{pb}}^*$ in Fig.~\ref{fig:mean-field phase diagram}. In between $v_{\mathrm{pb}}$ and $v_{\mathrm{pb}}^*$, a solution of gapless superfluid states exists within the mean-field approximation.
We also plot the sound velocity $c_0=\sqrt{(N/m)(\partial \mu /\partial N)}$, which corresponds to the slope of the AB phonon mode at the long-wavelength limit for $v=0$~\cite{combescot.pra.06}. As reported in Ref.~\cite{combescot.pra.06}, the curves of $v_{\mathrm{pb}}$ and $c_0$ cross each other near unitarity [$(1/k_{\mathrm{F}}a_s) _0\approx 0.075$]. For $1/k_{\mathrm{F}}a_s>(1/k_{\mathrm{F}}a_s) _0$, the collective phonon excitations occur at $v=c_0$ according to the Landau criterion~(\ref{eq:landau criterion}) before the quasiparticle-pair excitations. Therefore, the critical velocity for the breakdown of superfluidity is given by the sound velocity $c_0$~\cite{combescot.pra.06}. In contrast, for $1/k_{\mathrm{F}}a_s<(1/k_{\mathrm{F}}a_s) _0$, the quasiparticle-pair (Cooper pair-breaking) excitations start to be created at $v=v_{\mathrm{pb}}>c_0$, and the gapless superfluid state can exist in the region $v_{\mathrm{pb}}<v<v_{\mathrm{pb}}^*$ (see Fig.~\ref{fig:mean-field phase diagram}) according to our mean-field analysis in which a finite center-of-mass momentum of the Cooper pairs is considered.

\section{\label{sec3}linear stability analysis}

In this section, we analyze the linear stability of the mean-field solutions, especially focusing on the flow-induced gapless superfluid state. To this end, we study the collective-mode excitations applying the GRPA~\cite{engelbrecht.prb.97,combescot.pra.06,cote.prb.93,yunomae.pra.09,minguzzi.epjd.01,bruun.prl.01,ohashi.jpsj.97,ohashi.pra.03,taylor.pra.06,zhang.pra.11,hu.fp.12}. The GRPA analysis allows us to calculate the dynamic structure factor, which can be measured experimentally with the Bragg spectroscopy~\cite{veeravalli.prl.08,kuhnle.prl.10,hoinka.prl.13}.

\subsection{\label{subsec:grpa}Generalized random-phase approximation}

To discuss a superfluid system flowing with constant velocity, it is convenient to define a phase-twisted fermion operator~\cite{yunomae.pra.09}
\begin{align*}
\hat{\tilde{\psi }}^\dagger _\sigma (\mathbf{r})=\hat{\psi }^\dagger _\sigma (\mathbf{r})e^{im\mathbf{v}\cdot \mathbf{r}},
\end{align*}
and rewrite the mean-field Hamiltonian (\ref{eq:mean-field hamiltonian in nambu-gor'kov formalism}) as
\begin{align}\label{eq:mean-field hamiltonian in real space2}
\hat{H}_{\mathrm{MF}}=&\sum_{\sigma}\int d\mathbf{r}\,\hat{\tilde{\psi }}_{\sigma}^\dagger (\mathbf{r})\left[-\dfrac{(\nabla +im\mathbf{v})^2}{2m}-\mu \right]\hat{\tilde{\psi }}_{\sigma}(\mathbf{r}) \nonumber \\
&\hspace{15pt}-\int d\mathbf{r}\,(\hat{\tilde{\psi}}_{\uparrow}^\dagger (\mathbf{r})\hat{\tilde{\psi}}_{\downarrow}^\dagger (\mathbf{r})\Delta +\mathrm{H.c.})-\dfrac{\abs{\Delta }^2}{g}V
\end{align}
in real space. We discuss the linear response of the system to a time-dependent perturbation $U(\mathbf{r},t)$ that couples to the density $\hat{n}(\mathbf{r})=\sum_\sigma\hat{n}_\sigma(\mathbf{r})$. Here, $\hat{n}_\sigma(\mathbf{r})\equiv \hat{\psi }^{\dagger }_{\sigma }(\mathbf{r})\hat{\psi }_{\sigma }(\mathbf{r})=\hat{\tilde{\psi }}^{\dagger }_{\sigma }(\mathbf{r})\hat{\tilde{\psi }}_{\sigma }(\mathbf{r})$. The density response function matrix $\bm{\varPi}(\mathbf{r},t)$ to the external field
\begin{align*}
\hat{H}_{\mathrm{ext}}(t)=\int d\mathbf{r}\,\hat{n}(\mathbf{r})U(\mathbf{r},t)
\end{align*}
is defined by
\begin{align}\label{eq:local fluctuations of densities}
\delta \bm{\rho }(\mathbf{r},t)=\int d\mathbf{r}'\int dt'\,\bm{\varPi}(\mathbf{r}-\mathbf{r}',t-t')U(\mathbf{r}',t').
\end{align}
Here the matrix $\delta \bm{\rho }(\mathbf{r},t)$ represents the local fluctuations of the density matrix~\cite{ohashi.pra.03,combescot.pra.06}
\begin{align}\label{eq:generalized density matrix}
\hat{\bm{\rho }}(\mathbf{r})=
\begin{pmatrix}
\hat{n}_{\uparrow}(\mathbf{r}) & \hat{m}(\mathbf{r})\\[2pt]
\hat{m}^\dagger (\mathbf{r}) & -\hat{n}_{\downarrow}(\mathbf{r})
\end{pmatrix},
\end{align}
where $\hat{m}(\mathbf{r})\equiv \hat{\psi }_{\downarrow}(\mathbf{r})\hat{\psi }_{\uparrow }(\mathbf{r})e^{-2im\mathbf{v}\cdot \mathbf{r}}=\hat{\tilde{\psi }}_{\downarrow}(\mathbf{r})\hat{\tilde{\psi }}_{\uparrow }(\mathbf{r})$ is the anomalous density operator. 

Let us calculate $\bm{\varPi}(\mathbf{r},t)$ within the GRPA.  Since the order parameter $\Delta$ can be described in terms of $\hat{m}(\mathbf{r})$ as $\Delta =-g\langle \hat{m}(\mathbf{r}) \rangle$, the fluctuations around the mean-field $\Delta$ in Eq.~(\ref{eq:mean-field hamiltonian in real space2}) give rise to the effective potential
\begin{align*}
\delta \hat{H}_{\mathrm{int}}(t)=g\int d\mathbf{r}\,(\hat{\rho }_{21} (\mathbf{r})\delta \rho_{12}(\mathbf{r},t)+\hat{\rho }_{12} (\mathbf{r})\delta \rho_{21}(\mathbf{r},t)), 
\end{align*}
which acts like an external field that couples to the anomalous density. Thus the total external field including the fluctuation effects of the mean field is given by $\hat{H}^{\rm eff}_{\mathrm{ext}}(t)=\hat{H}_{\mathrm{ext}}(t)+\delta \hat{H}_{\mathrm{int}}(t)$. The linear response theory with respect to the perturbation $\hat{H}^{\rm eff}_{\mathrm{ext}}(t)$ leads to
\begin{align}\label{eq:local fluctuations of densities2}
\delta \bm{\rho }(\mathbf{r},t)=&\int d\mathbf{r}'\int dt'\,\bm{\varPi}^0(\mathbf{r}-\mathbf{r}',t-t')U(\mathbf{r}',t') \nonumber \\
&+g\int d\mathbf{r}'\int dt'\,\bm{\Chi }^0(\mathbf{r}-\mathbf{r}',t-t')^\dagger \delta \rho _{12}(\mathbf{r}',t') \nonumber \\
&+g\int d\mathbf{r}'\int dt'\,\bm{\Chi }^0(\mathbf{r}-\mathbf{r}',t-t')\delta \rho _{21}(\mathbf{r}',t')
\end{align}
with
\begin{align*}
\bm{\varPi }^{0}(\mathbf{r}-\mathbf{r}',t-t')&=-i\langle [\, \hat{\bm{\rho }}(\mathbf{r},t),\Tr[ \bm{\sigma }_3\hat{\bm{\rho }}(\mathbf{r}',t')]\,] \rangle\varTheta(t-t'),\\
\bm{\Chi }^{0}(\mathbf{r}-\mathbf{r}',t-t')&=-i\langle [\, \hat{\bm{\rho }}(\mathbf{r},t), \hat{\rho }_{12}(\mathbf{r}',t')] \rangle\varTheta(t-t').
\end{align*}
Here, $\bm{\sigma }_3$ is the Pauli matrix
\begin{align*}
\bm{\sigma }_3=\begin{pmatrix}
1&0\\[2pt]
0& -1
\end{pmatrix},
\end{align*}
and $\hat{\bm{\rho }}(\mathbf{r},t)=e^{i\hat{H}_{\mathrm{MF}}t}\hat{\bm{\rho }}(\mathbf{r})e^{-i\hat{H}_{\mathrm{MF}}t}$ is the density matrix operator in the Heisenberg picture. Comparing Eqs.~(\ref{eq:local fluctuations of densities}) and (\ref{eq:local fluctuations of densities2}), we find the following self-consistent GRPA equation:
\begin{multline}\label{eq:GRPA equation}
\bm{\varPi }(\mathbf{q},\omega )=\bm{\varPi }^0(\mathbf{q},\omega ) +g\bm{\Chi }^0(\mathbf{q},\omega )\varPi _{21}(\mathbf{q},\omega ) \\
+g\bm{\Chi }^0(-\mathbf{q},-\omega )^\dagger \varPi _{12}(\mathbf{q},\omega ).
\end{multline}
Thus the off-diagonal elements of $\bm{\varPi }(\mathbf{q},\omega )$ are obtained by solving the linear matrix equation
\begin{align}\label{eq:off-diagonal elements of the GRPA equation}
\bm{\varGamma}(\mathbf{q},\omega )
\begin{pmatrix}
\varPi _{12}(\mathbf{q},\omega ) \\[2pt]
\varPi _{21}(\mathbf{q},\omega ) 
\end{pmatrix}=\begin{pmatrix}
\varPi _{12}^0(\mathbf{q},\omega ) \\[2pt]
\varPi _{21}^0(\mathbf{q},\omega ) 
\end{pmatrix}
\end{align}
with
\begin{align*}
\bm{\varGamma}(\mathbf{q},\omega )=\begin{pmatrix}
1-g\Chi _{21}^0(-\mathbf{q},-\omega )^* & -g\Chi _{12}^0(\mathbf{q},\omega )\\[2pt]
-g\Chi _{12}^0(-\mathbf{q},-\omega )^* & 1-g\Chi _{21}^0(\mathbf{q},\omega )
\end{pmatrix}.
\end{align*}
Substituting the solution of Eq.~(\ref{eq:off-diagonal elements of the GRPA equation}) in Eq.~(\ref{eq:GRPA equation}), we also obtain the diagonal elements of $\bm{\varPi }(\mathbf{q},\omega )$. According to the fluctuation-dissipation theorem, the dynamic structure factor can be calculated through $S(\mathbf{q},\omega )=-\im[\varPi (\mathbf{q},\omega )]/\pi $ at zero temperature from the density-density response function $\varPi (\mathbf{q},\omega )=\Tr [\bm{\sigma }_3\bm{\varPi }(\mathbf{q},\omega )]$.

The response function matrices  $\bm{\varPi }^0(\mathbf{q},\omega )$ and $\bm{\Chi }^0(\mathbf{q},\omega )$ at the simple mean-field level are given by
\begin{align}\label{eq:PiandChi}
\begin{split}
\bm{\varPi }^0(\mathbf{q},i\Omega _m)&=\bm{G}\bm{\sigma }_3\bm{G},\\
\bm{\Chi }^0(\mathbf{q},i\Omega _m)&=\begin{pmatrix}G_{12}\\[1pt] G_{22} \end{pmatrix}\begin{pmatrix} G_{11}& G_{12} \end{pmatrix}.
\end{split}
\end{align}
Here, $\Omega _m=2\pi m/\beta $ is the bosonic Matsubara frequency with $m$ being an integer and we use the notation
\begin{align}\label{eq:definition of GG}
&G_{\alpha \beta }G_{\gamma \delta } \nonumber \\
&\hspace{10pt}=\dfrac{1}{\beta V}\sum_{\mathbf{k},\omega _n}G_{\alpha \beta }(\mathbf{k}+\mathbf{q},i\omega _n+i\Omega_m)G_{\gamma \delta }(\mathbf{k},i\omega _n). 
\end{align}
Within the mean-field theory of Sec.~\ref{sec2}, the single-particle Green's function $\bm{G}(\mathbf{k},i\omega _n)$ in the Nambu representation has the form~\cite{cote.prb.93}
\begin{align*}
\bm{G}(\mathbf{k},i\omega _n)=\dfrac{\bm{C}^+(\mathbf{k})}{i\omega _n-\omega _{\mathbf{k}}}+\dfrac{\bm{C}^-(\mathbf{k})}{i\omega _n+\omega _{-\mathbf{k}}},
\end{align*}
where $\omega _n=\pi (2n+1)/\beta $ is the fermionic Matsubara frequency with $n$ being an integer. The coefficient matrices $\bm{C}^\pm (\mathbf{k})$ are given by
\begin{align*}
\bm{C}^+(\mathbf{k})=
\begin{pmatrix}
{u_{\mathbf{k}}}{}^2 & u_{\mathbf{k}}v_{\mathbf{k}}\\[2pt]
u_{\mathbf{k}}v_{\mathbf{k}} & {v_{\mathbf{k}}}{}^2
\end{pmatrix},
\qquad 
\bm{C}^-(\mathbf{k})=\bm{1}-\bm{C}^+(\mathbf{k})
\end{align*}
where $u_{\mathbf{k}}$ and $v_{\mathbf{k}}$ are the Bogoliubov coefficients in Eq.~(\ref{eq:bogoliubov transformation}), and $\bm{1}$ is the $2\times 2$ unit matrix. Carrying out the summation over the fermionic Matsubara frequency $\omega _n$ in Eq.~(\ref{eq:definition of GG}) in the usual manner, we obtain
\begin{align}\label{eq:GG}
&G_{\alpha \beta }G_{\gamma \delta } \nonumber \\
=&\,\dfrac{1}{V}\sum_{\mathbf{k}}C_{\alpha \beta}^+(\mathbf{k}+\mathbf{q})C_{\gamma \delta }^-(\mathbf{k})\dfrac{1-f(\omega _{\mathbf{k}+\mathbf{q}})-f(\omega _{-\mathbf{k}})}{i\Omega _m-\omega _{\mathbf{k}+\mathbf{q}}-\omega _{-\mathbf{k}}} \nonumber \\
&+\dfrac{1}{V}\sum_{\mathbf{k}}\,[C_{\alpha \beta}^+(\mathbf{k}+\mathbf{q})C_{\gamma \delta }^+(\mathbf{k})+C_{\alpha \beta}^-(\mathbf{k})C_{\gamma \delta }^-(\mathbf{k}+\mathbf{q})] \nonumber \\[-5pt]
&\hspace{130pt}\times\dfrac{-f(\omega _{\mathbf{k}+\mathbf{q}})+f(\omega _{\mathbf{k}})}{i\Omega _m-\omega _{\mathbf{k}+\mathbf{q}}+\omega _{\mathbf{k}}} \nonumber \\[4pt]
&+\dfrac{1}{V}\sum_{\mathbf{k}}C_{\alpha \beta}^-(\mathbf{k}+\mathbf{q})C_{\gamma \delta }^+(\mathbf{k})\dfrac{f(\omega _{-\mathbf{k}-\mathbf{q}})+f(\omega _{\mathbf{k}})-1}{i\Omega _m+\omega _{-\mathbf{k}-\mathbf{q}}+\omega _{\mathbf{k}}},
\end{align}
where $f(\epsilon )=1/(1+e^{\beta \epsilon})$ is the Fermi-Dirac distribution function with $\beta =1/T$ being the inverse temperature. At zero temperature, $f(\epsilon )$ is replaced by the Heaviside step function $\varTheta (-\epsilon)$. The analytic continuation $i\Omega _m\to \omega +i\delta $ (where $\delta $ is a positive infinitesimal) gives $\bm{\varPi }^0(\mathbf{q},\omega )$ and $\bm{\Chi }^0(\mathbf{q},\omega )$ from Eqs.~(\ref{eq:PiandChi}) and (\ref{eq:GG}). Now one can calculate the dynamic structure factor $S(\mathbf{q},\omega )$ from Eqs.~(\ref{eq:GRPA equation})-(\ref{eq:GG}).

\subsection{\label{subsec:ce}Excitation spectrum}

\begin{figure}
\includegraphics[width=8.2cm]{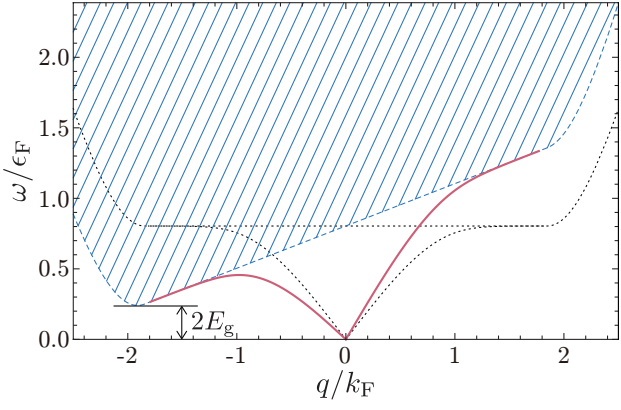}
\caption{\label{fig:excitation spectra(gapped)}
(Color online) Excitation spectrum for $1/k_{\mathrm{F}}a_s=-0.5$ and $v=0.15v_{\mathrm{F}}$. The dashed area corresponds to the continuum of the quasiparticle-pair excitations and the red-solid line shows the dispersion relation of the gapless AB mode. The boundary of the continuum and the AB mode dispersion for $v=0$~\cite{combescot.pra.06} are also shown by the dotted lines.  A positive (negative) value of $q$ represents the magnitude of the quasiparticle momentum $\mathbf{q}$ which is in the same (opposite) direction as the flow.}
\end{figure}

The GRPA excitation spectrum $\omega(\mathbf{q})$ is given as poles of the structure factor $S(\mathbf{q},\omega )$ or, equivalently, of the response function $\bm{\varPi} (\mathbf{q},\omega )$. In what follows, we restrict our attention to the excitations with momentum $\mathbf{q}\parallel \mathbf{v}$, which is crucial for the instability caused by superflow. We use a positive (negative) value of $q$ when $\mathbf{q}$ and $\mathbf{v}$ is in the same (opposite) direction. As shown in Fig.~\ref{fig:excitation spectra(gapped)}, the excitation spectrum of the standard (gapped) superfluid state for $v<v_{\mathrm{pb}}$ consists of two different types of excitations; one is the quasiparticle-pair continuum with the energy gap $2E_{\mathrm{g}}$ and the other is a gapless collective-mode dispersion~\cite{combescot.pra.06}. For $v<v_{\mathrm{pb}}$, introducing superflow causes only the tilting of the excitation spectrum in the opposite direction of the superflow. 
The continuum comes from the poles of the summands of the first sum in Eq.~(\ref{eq:GG}), which correspond to the quasiparticle-pair excitations with the energy $\omega _{\mathbf{k}+\mathbf{q}}+\omega _{-\mathbf{k}}$. Minimizing $\omega _{\mathbf{k}+\mathbf{q}}+\omega _{-\mathbf{k}}$ for given $\mathbf{q}$ with respect to $\mathbf{k}$, we obtain the lower bound of the continuum
\begin{align*}
&\omega ^{\text{pp}}_{-}(q)\\
&\hspace{10pt}=\begin{cases}
2\Delta +vq & (\text{$\mu>0$ and $|q|\le 2\sqrt{2m\mu -(mv)^2}$}), \\
\omega _{q/2} & (\text{otherwise}).
\end{cases}
\end{align*}
The second sum in Eq.~(\ref{eq:GG}) does not contribute to the excitation spectra at zero temperature for $v<v_{\mathrm{pb}}$ because $\omega _{\mathbf{k}}$ is always positive for any value of ${\mathbf{k}}$. The collective-mode dispersion, known as the AB mode, can be obtained by solving the equation $\det \bm{\varGamma}(\mathbf{q},\omega )=0$ [see Eq.~(\ref{eq:off-diagonal elements of the GRPA equation})] together with the gap equation (\ref{eq:gap equation}). The collective mode exhibits a gapless linear dispersion in the long-wavelength region: 
\begin{align}\label{eq:collective mode dispersion (gapped)}
\omega (\mathbf{q})\approx c_0q+\mathbf{q}\cdot \mathbf{v}. 
\end{align}
This agrees with the hydrodynamic description of the phonon dispersion for a moving condensate~\cite{zaremba.pra.98,machholm.pra.03,taylor.pra.03}, i.e., 
\begin{align}\label{eq:hydrodynamic}
\omega (\mathbf{q})\approx \sqrt{\bar{\mathcal{E}}_{N,N}\bar{\mathcal{E}}_{v_i,v_j}q_iq_j/m^2}+\bar{\mathcal{E}}_{N,v_i}q_i/m, 
\end{align}
where $\bar{\mathcal{E}}=\mathcal{E}_0+\mu N$, $\bar{\mathcal{E}}_{N,N}=\partial ^2\bar{\mathcal{E}}/\partial N^2$ and $\bar{\mathcal{E}}_{v_i,v_j}=\partial ^2\bar{\mathcal{E}}/\partial v_i \partial v_j$. Here we used the Einstein convention for summation over repeated Cartesian indices $i$ and $j$.

The amplitude and phase fluctuations of the superfluid order parameter can be described by $\lambda (\mathbf{q},\omega )=[\varPi _{12}(\mathbf{q},\omega )+\varPi _{21}(\mathbf{q},\omega )]/\sqrt{2}$ and $\theta (\mathbf{q},\omega )=[\varPi _{12}(\mathbf{q},\omega )-\varPi _{21}(\mathbf{q},\omega )]/\sqrt{2}$, respectively~\cite{engelbrecht.prb.97}. We also define $\lambda ^0(\mathbf{q},\omega )$ and $\theta ^0(\mathbf{q},\omega )$ for $\varPi^0_{12}(\mathbf{q},\omega )$ and $\varPi^0_{21}(\mathbf{q},\omega )$ in the same way. In terms of these quantities, Eq.~(\ref{eq:off-diagonal elements of the GRPA equation}) becomes
\begin{align}\label{eq:off-diagonal elements of the GRPA equation2}
\begin{pmatrix}
A(\mathbf{q},\omega ) & M(\mathbf{q},\omega ) \\[2pt]
M(\mathbf{q},\omega ) & P(\mathbf{q},\omega )
\end{pmatrix}
\begin{pmatrix}
\lambda (\mathbf{q},\omega ) \\[2pt]
\theta (\mathbf{q},\omega ) 
\end{pmatrix}=
\begin{pmatrix}
\lambda ^0(\mathbf{q},\omega ) \\[2pt]
\theta ^0(\mathbf{q},\omega ) 
\end{pmatrix},
\end{align}
where $A(\mathbf{q},\omega )=1-g\Chi_{21}^{+}(\mathbf{q},\omega )-g\Chi_{12}^0(\mathbf{q},\omega )$, $P(\mathbf{q},\omega )=1-g\Chi _{21}^{+}(\mathbf{q},\omega )+g\Chi _{12}^0(\mathbf{q},\omega )$, and $M(\mathbf{q},\omega )=g\Chi_{21}^{-}(\mathbf{q},\omega )$ with $\Chi _{21}^{\pm}(\mathbf{q},\omega )=[\Chi _{21}^0(\mathbf{q},\omega )\pm \Chi _{21}^0(-\mathbf{q},-\omega )^*]/2$. Here we used the relation $\Chi_{12}^0(\mathbf{q},\omega )=\Chi_{12}^0(-\mathbf{q},-\omega )^*$. The off-diagonal element $M(\mathbf{q},\omega )$ represents the coupling between the amplitude and phase fluctuations.  In the gapped superfluid state for $v<v_{\mathrm{pb}}$, the amplitude and phase modes can be decoupled when the Doppler-shifted frequency $\tilde{\omega }\equiv\omega -\mathbf{q}\cdot \mathbf{v}$~\cite{taylor.pra.06} is zero, i.e., $M(\mathbf{q},\mathbf{q}\cdot \mathbf{v})=0$. Moreover, the fact $P(\mathbf{0},0)=0$ indicates that the long-wavelength and low-energy excitation of the gapless AB collective mode in Fig.~\ref{fig:excitation spectra(gapped)} arises from the phase fluctuations. This is indeed a Nambu-Goldstone mode responsible for {\rm U(1)} symmetry breaking~\cite{engelbrecht.prb.97}. The other branch of collective motion, arising from the amplitude fluctuations of the order parameter, has a gap $2\Delta$ at $\mathbf{q}=\mathbf{0}$, i.e., $A(\mathbf{0},2\Delta)=0$. The dispersion of the amplitude mode is not visible in Fig.~\ref{fig:excitation spectra(gapped)} since it merges into the quasiparticle-pair continuum in contrast to exceptional cases such as superfluids in the honeycomb lattice~\cite{tsuchiya.prb.13}.

\begin{figure}
\includegraphics[width=8.4cm]{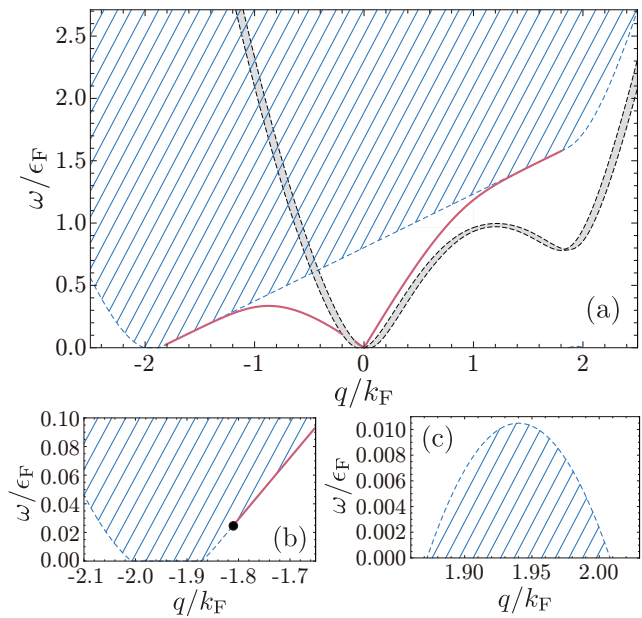}
\caption{\label{fig:excitation spectra(gapless)}
(Color online) Excitation spectrum for $1/k_{\mathrm{F}}a_s=-0.5$ and $v=0.215v_{\mathrm{F}}$. (b) and (c) are enlarged views of (a) near $q=\pm 2k_{\mathrm{F}}$ and low $\omega $. The red-solid line shows the dispersion relation of the collective mode. The striped and gray-shaded areas correspond to the region where the quasiparticle-quasiparticle (or quasihole-quasihole) and quasiparticle-quasihole excitations are possible, respectively. The dashed lines represent the boundaries of those continua. A positive (negative) value of $q$ represents the magnitude of the quasiparticle momentum $\mathbf{q}$ which is in the same (opposite) direction as the flow.}
\end{figure}

When the superflow velocity $v$ exceeds $v_{\mathrm{pb}}$ given by Eq.~(\ref{eq:pair-breaking velocity}), the spontaneous quasiparticle-pair creations (the Cooper pair-breaking excitations) occur and the negative energy states in the Bogoliubov quasiparticle bands are occupied. As a result, the feature of the excitation spectrum is qualitatively changed. In Fig.~\ref{fig:excitation spectra(gapless)}, we show the GRPA excitation spectrum of the flow-induced gapless superfluid state for $1/k_{\mathrm{F}}a_s=-0.5$ and $v=0.215v_{\mathrm{F}}>v_{\mathrm{pb}}$.
One can see the appearance of an additional narrow continuum (the gray-shaded area), which comes from the poles of the summands of the second sum in Eq.~(\ref{eq:GG}). This excitation corresponds to the simultaneous creations of a quasiparticle with the energy $\omega _{\mathbf{k}+\mathbf{q}}>0$ and a quasihole with $\omega _{\mathbf{k}}<0$. In other words, the excitation process can be seen as the intra-band scattering of a quasiparticle $\hat{\alpha }_{\mathbf{k}\lambda }$ from an occupied state [the red-dashed curve in Fig.~\ref{fig:quasiparticle dispersions}(d)] to an empty state (the green-solid curve). In Fig.~\ref{fig:excitation spectra(gapless)}(a), we determine the upper and lower boundaries of the ``quasiparticle-quasihole'' continuum, which are obtained by maximizing and minimizing the quasiparticle-quasihole creation energy $\omega _{\mathbf{k}+\mathbf{q}}-\omega _{\mathbf{k}}$ $(\omega _{\mathbf{k}+\mathbf{q}}>0,\,\omega _{\mathbf{k}}<0)$ for given $\mathbf{q}$ with respect to $\mathbf{k}$. Furthermore, the gap of the quasiparticle-quasiparticle continuum, $2E_{\mathrm{g}}$, vanishes and the curve of the lower bound $\omega ^{\text{pp}}_{-}(q)$ is truncated at $\omega= 0$ around $q=-2k_{\rm F}$ [see Fig.~\ref{fig:excitation spectra(gapless)}(b)]. Instead, the third sum in Eq.~(\ref{eq:GG}) causes the ``quasihole-quasihole'' continuum around $q=2k_{\rm F}$ [Fig.~\ref{fig:excitation spectra(gapless)}(c)].

The fact that a part of the Bogoliubov quasiparticle bands are filled also strongly affects the behavior of the collective-mode dispersion in the gapless superfluid state. In the next subsection, we will discuss the details and find that the collective-mode excitation can cause instability of the flow-induced gapless superfluid state.

\subsection{\label{subsec:di}Dynamical instability}

\begin{figure}
\includegraphics[width=8.4cm]{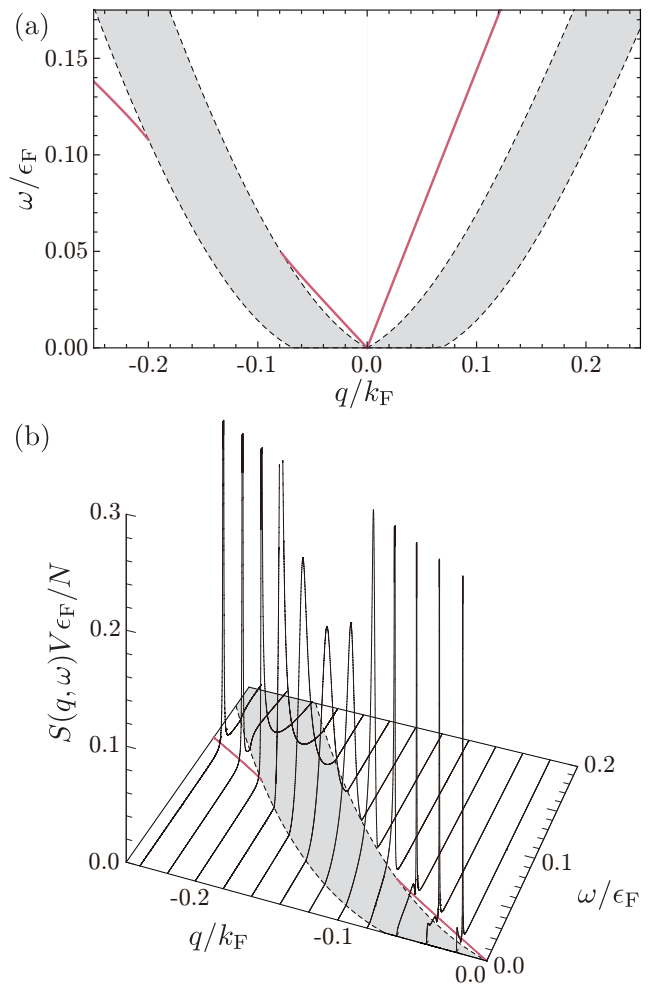}
\caption{\label{fig:excitation spectra and dynamic structure factor1}
(Color online) (a) Excitation spectrum in the long-wavelength region for $1/k_{\mathrm{F}}a_s=-0.5$ and $v=0.215v_{\mathrm{F}}$. (b) Dynamic structure factor in the left half of (a). In calculating the dynamic structure factor, we use a broadening factor of $\delta =10^{-4}\epsilon_{\mathrm{F}}$.}
\end{figure}

In Fig.~\ref{fig:excitation spectra and dynamic structure factor1}(a), we show an enlarged view of the low-energy and long-wavelength region in Fig.~\ref{fig:excitation spectra(gapless)}(a).
\begin{figure}
\includegraphics[width=8.4cm]{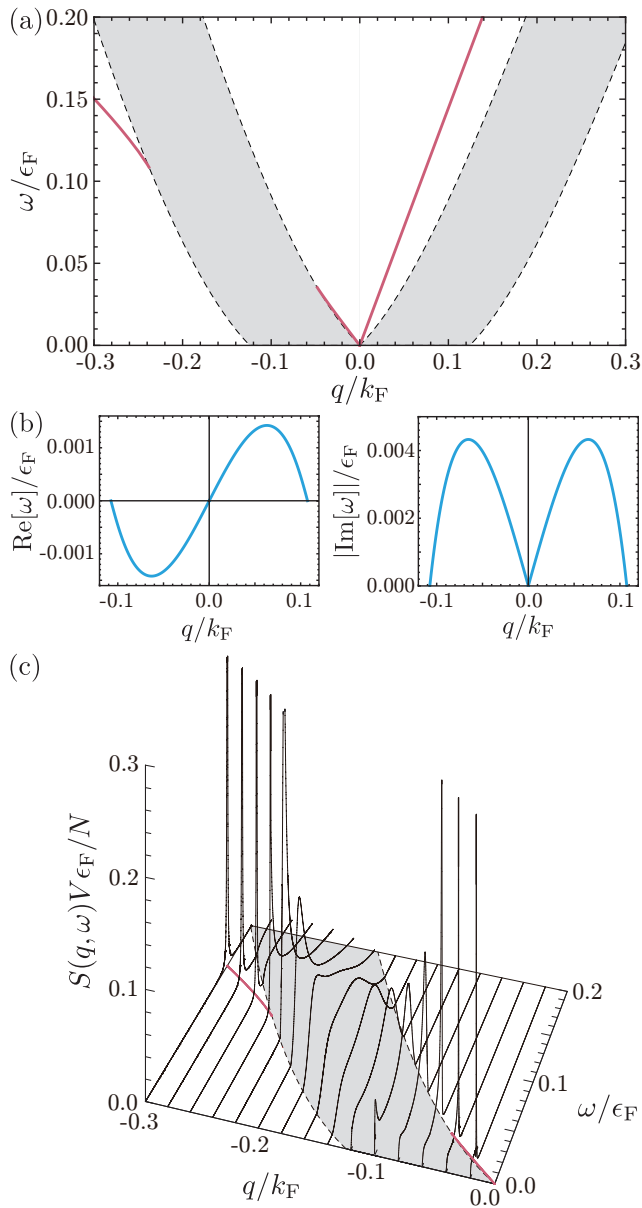}
\caption{\label{fig:excitation spectra and dynamic structure factor}
(Color online) (a) Excitation spectrum and (b) excitation energy of the second collective mode in the long-wavelength region for $1/k_{\mathrm{F}}a_s=-0.5$ and $v=0.22v_{\mathrm{F}}$. (c) Dynamic structure factor in the left half of (a). We use a broadening factor of $\delta =10^{-4}\epsilon_{\mathrm{F}}$.}
\end{figure}
We numerically find that the slope of the collective mode at $|\mathbf{q}|\approx 0$ disagrees with the standard hydrodynamic expression given by Eq.~(\ref{eq:hydrodynamic}) in contrast to the case of the gapped superfluid. Moreover, the amplitude and phase fluctuations are decoupled only when $\tilde{\omega }=0$ AND $\mathbf{q}=0$ since $M(\mathbf{q},\mathbf{q}\cdot \mathbf{v})\neq 0$ for $\abs{\mathbf{q}} \neq 0$. As shown in Fig.~\ref{fig:excitation spectra and dynamic structure factor1}(a), the curve of the gapless collective mode passes through the quasiparticle-quasihole continuum band. In order to see the merging of the collective-mode and continuum excitations in detail, we calculate the dynamic structure factor $S(\mathbf{q},\omega)$. As shown in Fig.~\ref{fig:excitation spectra and dynamic structure factor1}(b), the sharp delta-function peak is broadened inside the quasiparticle-quasihole continuum, which indicates damping of the collective mode. Moreover, one can also notice that a small subpeak exists in $S(\mathbf{q},\omega)$ for very small values of $|\mathbf{q}|$ in addition to the sharp delta-function peak. This may indicate the appearance of a second gapless collective mode, although it is strongly damped due to the merging with the continuum. Those characteristic features of the GRPA excitation spectrum for the gapless superfluid state should be intimately related to the fact that the Galilean symmetry of the system is broken due to the spontaneous quasiparticle-pair excitations.

\begin{figure}
\includegraphics[width=8.4cm]{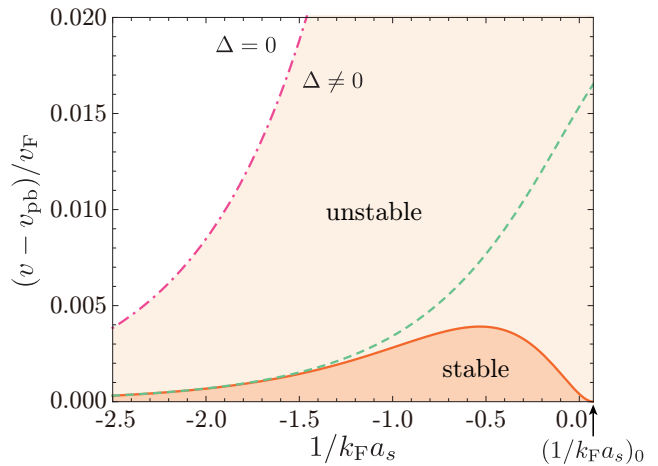}
\caption{\label{fig:stability phase diagram}
(Color online)
The $1/k_{\mathrm{F}}a_s$ dependences of $v_{\mathrm{pb}}^*-v_{\mathrm{pb}}$ (solid line), $v_{\mathrm{di}}-v_{\mathrm{pb}}$ (dot-dashed line) and $\bar{v}-v_{\mathrm{pb}}$ (dashed line) in units of the Fermi velocity $v_{\mathrm{F}}$. The stable gapless superfluid state can be formed in the region below the solid line. Above the dot-dashed line, the superfluid order parameter $\Delta $ vanishes.}
\end{figure}

When the superflow velocity further increases and exceeds a certain critical value, the second collective mode is more clearly obtained as a solution of the GRPA equation for the collective mode, $\det \bm{\varGamma}(\mathbf{q},\omega )=0$. We show the excitation spectrum and the second solution of the collective mode for $1/k_{\mathrm{F}}a_s=-0.5$ and $v=0.22v_{\mathrm{F}}$ in Fig.~\ref{fig:excitation spectra and dynamic structure factor}(a) and \ref{fig:excitation spectra and dynamic structure factor}(b), respectively. As shown in Fig.~\ref{fig:excitation spectra and dynamic structure factor}(b), the long-wavelength excitation energy of the second collective mode has a nonzero imaginary part. The imaginary part vanishes for $q\lesssim -0.1k_{\mathrm{F}}$, and the second collective mode merges into the continuum. Although the damped collective mode is hardly visible, one can still notice a small subpeak in the structure factor $S(q,\omega)$ around $q=-0.1k_{\mathrm{F}}$ [Fig.~\ref{fig:excitation spectra and dynamic structure factor}(c)]. The appearance of excitations with a nonzero imaginary part can be seen as a signal of dynamical instability of the system since it causes an exponential growth of perturbations in time~\cite{wu.pra.01,burkov.prl.08}. We calculate the velocity $v_{\mathrm{di}}$ at which a long-wavelength excitation with $\im [\omega]\neq 0$ begins to appear, and determine the stability phase diagram for $v\geq v_{\mathrm{pb}}$ in Fig.~\ref{fig:stability phase diagram}.
In Ref.~\cite{combescot.pra.06}, the critical velocity on the BCS side has been determined by the onset of fermionic pair-breaking excitations, $v_{\mathrm{pb}}$. However, in the narrow region of Fig.~\ref{fig:stability phase diagram} below the curve of $v_{\mathrm{di}}-v_{\mathrm{pb}}$, the gapless superfluid state can be still dynamically stable. This means that the consideration of gapless superfluid state leads to an increase in the critical velocity of superflow on the BCS side, although the difference from the pair-breaking onset velocity $v_{\mathrm{pb}}$ is very small. Our result suggests that although the pair-breaking excitations do not immediately cause the instability of superflow, the dynamical destabilization occurs due to the second collective mode with a nonzero imaginary part in the gapless superfluid state. Note that $v_{\mathrm{di}}$, $v_{\mathrm{pb}}$, and $c_0$ coincide at $1/k_{\mathrm{F}}a_s=(1/k_{\mathrm{F}}a_s)_0$. Thus the critical velocity is given by $v=v_{\mathrm{di}}$ for $1/k_{\mathrm{F}}a_s<(1/k_{\mathrm{F}}a_s)_0$ and $v=c_0$ for $1/k_{\mathrm{F}}a_s>(1/k_{\mathrm{F}}a_s)_0$, and the point at which the instability mechanism switches in the BCS-BEC crossover does not change from that obtained in Ref.~\cite{combescot.pra.06}.

In usual gapped superfluids, the instability of the system has been also discussed by using the appearance of a negative value of a second derivative of the free energy as a criterion~\cite{watanabe.pra.09,watanabe.pra.11}. Especially, the phase stiffness $\propto \partial ^2 \bar{\mathcal{E}}/\partial v^2$ is useful under the existence of superflow since it is connected to the long-wavelength collective-mode excitations via Eq.~(\ref{eq:hydrodynamic}). We therefore calculate $\partial ^2 \bar{\mathcal{E}}/\partial v^2$ in the gapless superfluid state for $v>v_{\mathrm{pb}}$, in which Eq.~(\ref{eq:hydrodynamic}) is no longer valid. For $0 \le v \le v_{\mathrm{pb}}$, the quantity $m^2(\partial ^2 \bar{\mathcal{E}}/\partial v^2)^{-1}$ equals to the bare fermion mass $m$ regardless the value of $v$. If the velocity $v$ exceeds $v_{\mathrm{pb}}$, the {\it effective} mass $m^*=m^2(\partial ^2 \bar{\mathcal{E}}/\partial v^2)^{-1}$ deviates from $m$ and eventually diverges at a certain velocity $\bar{v}$. In other words, the sign of $\partial ^2 \bar{\mathcal{E}}/\partial v^2$ changes from positive to negative at $v=\bar{v}$. As shown in Fig.~\ref{fig:stability phase diagram}, the value of $\bar{v}$ does not coincide with $v_{\mathrm{di}}$ and a negative value of $\partial ^2 \bar{\mathcal{E}}/\partial v^2$ appears always after the dynamical instability occurs. This fact reminds us that the standard hydrodynamic relation Eq.~(\ref{eq:hydrodynamic}) is not satisfied in the GRPA for the gapless superfluid state. Note that $\bar{v}$ approaches asymptotically to $v_{\mathrm{di}}$ in the weak-coupling limit as can be seen in Fig.~\ref{fig:stability phase diagram}.

\section{\label{sec4}conclusion}

In conclusion, we have studied a superfluid Fermi gas flowing with constant velocity $v$ in three-dimensional free space, especially focusing on the BCS side of the BCS-BEC crossover. Within the mean-field theory, we found that the superfluid order parameter persists even after the gap in the Bogoliubov quasiparticle dispersion closes at $v=v_{\mathrm{pb}}$ due to the Doppler shift. In the flow-induced gapless superfluid state, the superfluid order parameter monotonically decreases with increasing the superflow velocity because of the quasiparticle-pair creations (the Cooper pair-breaking excitations) and eventually vanishes at $v=v_{\mathrm{pb}}^*$. 

For $v_{\mathrm{pb}}<v<v_{\mathrm{pb}}^*$, the effect of superflow cannot be understood by simply introducing a Doppler shift in excitation energies since the spontaneous quasiparticle-pair creation breaks the Galilean symmetry of the system. In this case, the critical velocity of superflow is not determined by the Landau criterion for the stationary state, which is a similar situation to superfluids on a lattice~\cite{ganesh.pra.09,yunomae.pra.09}. Therefore, using the GRPA with explicitly considering a nonzero center-of-mass momentum of Cooper pairs, we analyze the fluctuations around the mean-field solution and the linear stability of the flow-induced gapless superfluid state. In a stationary superfluid state, the GRPA energy spectrum consists of quasiparticle-pair continuum and a gapless collective-mode dispersion whose slope at the long-wavelength limit is the sound velocity $c_0$~\cite{combescot.pra.06}. The effect of superflow is nothing more than tilting of the energy spectrum up to $v=v_{\mathrm{pb}}$. However, the structure of the GRPA spectrum becomes qualitatively different for $v_{\mathrm{pb}}<v<v_{\mathrm{pb}}^*$. In the gapless superfluid state, the quasiparticle-quasihole and quasihole-quasihole continua appear in addition to the quasiparticle-pair continuum. Moreover, a second collective mode is found at the long-wavelength region. Although the second collective mode causes a dynamical instability slightly above $v=v_{\mathrm{pb}}$, the gapless superfluid state is expected to survive as a stable phase in a narrow region of $v>v_{\mathrm{pb}}$. 

Finally let us briefly mention other gapless superfluid states, such as the Sarma state~\cite{sarma.jpcs.63}, the Fulde-Ferrell-Larkin-Ovchinnikov (FFLO) state~\cite{fulde.pr.64,larkin.zetf.65}, and the breached-pairing state~\cite{liu.prl.03}. These exotic gapless superfluid states are predicted to appear, e.g., in two-component attractive Fermi gases with mismatched Fermi surfaces~\cite{iskin.prl.06,hu.pra.06}. By analogy with the flow-induced gapless superfluid states studied here, those exotic superfluid states might also suffer from dynamical instability due to the emergence of the second collective mode with a nonzero imaginary part in its long-wavelength excitation energy. Note that, although the superfluidity of population-imbalanced Fermi gases has been realized in ultracold atomic systems~\cite{zwierlein.sci.06,partridge.sci.06}, the exotic gapless superfluid states have not been identified experimentally yet. Our study suggests that the standard hydrodynamic approach cannot be applied to the gapless superfluid states. In this case, the appearance of the complex-frequency collective mode has no direct relation to the sign of a second derivative of the free energy, for instance, phase stiffness.

\begin{acknowledgments}
We would like to thank Shunji Tsuchiya and Ippei Danshita for useful discussions. We also acknowledge helpful comments of Keisuke Masuda and Susumu Kurihara. 
\end{acknowledgments}

\begin{appendix}

\section{\label{appendix} Derivation of Eq.~(\ref{eq:order parameter vanishes})}

Here we present a detailed derivation of Eq.~(\ref{eq:order parameter vanishes}). As seen in Fig.~\ref{fig:self-consistent solutions}(a), the order parameter vanishes continuously at $v=v_{\mathrm{pb}}^*$ since the superfluid-to-normal transition is of second order at the mean-field level. The second-order transition point $v_{\mathrm{pb}}^*$ is determined by solving the gap equation (\ref{eq:gap equation2}) along with the number equation (\ref{eq:number equation}). From the number equation (\ref{eq:number equation}), it is easily seen that the chemical potential is equal to the Fermi energy $\epsilon _{\mathrm{F}}$ for $\Delta =0$. Substituting $\Delta =0$ and $\mu =\epsilon _{\mathrm{F}}$, we obtain the following reduced gap equation:
\begin{align}\label{eq:second order transition}
-\dfrac{m}{4\pi a_s}&=\dfrac{1}{4\pi ^2}\int_0^\infty dk\int_0^1dt\dfrac{k^2}{\abs{\xi^{*+}_{\mathrm{pb}}}}\varTheta(\abs{\xi^{*+}_{\mathrm{pb}}}-kv_{\mathrm{pb}}^*t) \nonumber \\
&\hspace{100pt}-\dfrac{1}{4\pi ^2}\int_0^\infty dk\dfrac{k^2}{\epsilon_{\mathbf{k}}}.
\end{align}
Here we use the notation $\xi^{*+}_{\mathrm{pb}}=[k^2+(mv_{\mathrm{pb}}^*)^2-{k_{\mathrm{F}}}^2]/2m$, and $t$ is the cosine of the angle between $\mathbf{k}$ and $\mathbf{v}_{\mathrm{pb}}^*$. Using the relation $\varTheta(\abs{\xi^{*+}_{\mathrm{pb}}}-kv_{\mathrm{pb}}^*t)=\varTheta(\omega^{*-}_{\mathrm{pb}})\varTheta(\xi^{*+}_{\mathrm{pb}})+\varTheta(-\omega^{*+}_{\mathrm{pb}})\varTheta(-\xi^{*+}_{\mathrm{pb}})$, where $\omega ^{*\pm }_{\mathrm{pb}}=\xi^{*+}_{\mathrm{pb}}\pm kv_{\mathrm{pb}}^*t$, we carrying out the integration in Eq.~(\ref{eq:second order transition}) and obtain
\begin{align}\label{eq:second order transition1}
\dfrac{\pi }{2k_{\mathrm{F}}a_s}=1+2\sqrt{\left(\dfrac{v_{\mathrm{pb}}^*}{v_{\mathrm{F}}}\right)^2-1}\arctan\sqrt{\dfrac{v_{\mathrm{pb}}^*-v_{\mathrm{F}}}{v_{\mathrm{pb}}^*+v_{\mathrm{F}}}},
\end{align}
for $v_{\mathrm{pb}}^*>v_{\mathrm{F}}$ and
\begin{align}\label{eq:second order transition2}
\dfrac{\pi }{2k_{\mathrm{F}}a_s}=1-2\sqrt{1-\left(\dfrac{v_{\mathrm{pb}}^*}{v_{\mathrm{F}}}\right)^2}\arctanh\sqrt{\dfrac{v_{\mathrm{F}}-v_{\mathrm{pb}}^*}{v_{\mathrm{F}}+v_{\mathrm{pb}}^*}},
\end{align}
for $v_{\mathrm{pb}}^*<v_{\mathrm{F}}$, respectively. Finally, one can obtain Eq.~(\ref{eq:order parameter vanishes}) from Eqs.~(\ref{eq:second order transition1}) and (\ref{eq:second order transition2}) by introducing $\theta=\arccos (v_{\mathrm{F}}/v_{\mathrm{pb}}^*)$. Here we use the definition $\arccos x=i\arccosh x$ for $x>1$. From Eq.~(\ref{eq:second order transition2}), we find that $v_{\mathrm{pb}}^*/v_{\mathrm{F}}\sim 2e^{\frac{\pi }{2k_{\mathrm{F}}a_s}-1}$ in the limit of $1/k_{\mathrm{F}}a_s\to -\infty $. On the other hand, $v_{\mathrm{pb}}$ has the form $v_{\mathrm{pb}}\sim \sqrt{\mu_0/2m}\,(\Delta _0/\mu_0)\sim 4v_{\mathrm{F}}e^{\frac{\pi }{2k_{\mathrm{F}}a_s}-2}$ because $\mu _0\sim \epsilon_{\mathrm{F}}$ and $\Delta _0\sim 8e^{-2}\epsilon_{\mathrm{F}}e^{\frac{\pi }{2k_{\mathrm{F}}a_s}}$~\cite{engelbrecht.prb.97}, and therefore $v_{\mathrm{pb}}^*/v_{\mathrm{pb}}\to e/2$~\cite{zagoskin.tb.98,wei.prb.09} in the BCS limit.

\end{appendix}


\end{document}